\documentclass[pra,aps,graphicx,multicol,epsfig]{revtex4}
\usepackage{graphics}

\begin{document}

\title{Deviation from one-dimensionality in stationary properties and
collisional dynamics of matter-wave solitons}
\author{Lev Khaykovich$^{1}$ and Boris A. Malomed$^{2}$}
\affiliation{$^{1} $Department of Physics, Bar-Ilan University, Ramat-Gan, 52900 Israel,}
\affiliation{$^{2}$Department of Interdisciplinary Studies, School of Electrical
Engineering, Faculty of Engineering, Tel Aviv University, Tel Aviv 69978,
Israel}

\begin{abstract}
By means of analytical and numerical methods, we study how the residual
three-dimensionality affects dynamics of solitons in an attractive
Bose-Einstein condensate loaded into a cigar-shaped trap. Based on an
effective 1D Gross-Pitaevskii equation that includes an additional quintic
self-focusing term, generated by the tight transverse confinement, we find a
family of exact one-soliton solutions and demonstrate stability of the
entire family, despite the possibility of collapse in the 1D equation with
the quintic self-focusing nonlinearity. Simulating collisions between two
solitons in the same setting, we find a critical velocity, $V_{c}$, below
which merger of identical in-phase solitons is observed. Dependence of $%
V_{c} $ on the strength of the transverse confinement and number of atoms in
the solitons is predicted by means of the perturbation theory and
investigated in direct simulations. The simulations also demonstrate
symmetry breaking in collisions of identical solitons with a nonzero phase
difference. This effect is qualitatively explained by means of an analytical
approximation.
\end{abstract}

\maketitle

\section{Introduction}

It is well known that a trapped atomic Bose-Einstein condensate
(BEC) with attractive interactions is stable if the number of atoms
in it is below a critical value, above which collapse occurs
\cite{collapse}. Beneath the collapse threshold, the BEC can form
stable wave packets in a one-dimensional (1D) ``cigar-shaped" trap,
which is tightly confined in two (transverse) directions, and is
unbound along the longitudinal axis. In that case, the stability of
bright solitons is provided by balance between the quantum pressure,
alias matter-wave (MW) dispersion, and mean-field attraction. A single MW soliton \cite%
{Khaykovich02} and trains of interacting solitons \cite{Strecker02} have
been created in the cigar-shaped optical traps. However, while the trapping
geometry was nearly one-dimensional, the solitons themselves were far from
being 1D objects. In particular, in Ref. \cite{Khaykovich02}, a stable
soliton was only possible if its longitudinal size exceeded the transverse
size by no more than $20\%$ (note that the situation was affected by an
expulsive axial potential, unavoidable in the specific experimental setup).
Recently, it was shown that the proximity of the soliton to being a 3D
object strongly affects its properties, such as the character of its motion
\cite{Brand06} and interactions \cite{Adams06,Salerno}. In particular, it
was demonstrated that a moving soliton immersed in a cloud of thermal atoms
is subjected to a temperature-dependent friction force \cite{Brand06}. A
collision between two solitons, which are by themselves stable, in a
confined geometry may readily lead to collapse, if the total number of atoms
in the soliton pair exceeds the above-mentioned critical value, and the
phase difference between the solitons is (close to) zero \cite{Salerno}. The
significance of the effective dimensionality of MW solitary pulses is
further emphasized by the recent observation of formation of a set of
nearly-3D mutually repulsive MW solitons (with a phase shift of $\pi $
between them) as a result of incomplete collapse in an attractive BEC with
the number of atoms \emph{several times larger} than the critical value \cite%
{Cornish06, Adams06}.

In addition to being a profoundly important object of fundamental studies,
MW solitary waves are also natural candidates for applications, such as
high-precision atom interferometry and quantum-information processing. Thus,
a thorough understanding of deviations of their behavior from that of ideal
1D solitons is important in this respect too.

In this paper we report results of theoretical investigation of the shape of
stationary MW solitons and binary collisions between them in the quasi-1D
regime, with the aim to identify manifestations of nonsolitonic behavior due
to the residual multi-dimensionality. The effect of the tightly confined
transverse dimensions is taken into account through a perturbative \emph{%
self-focusing} quintic term added to the corresponding one-dimensional
Gross-Pitaevskii equation (GPE), as per Refs. \cite{Shlyap02} and \cite%
{Brand06}. In Section II, we introduce this extended GPE, find a family of
its exact one-soliton solutions, and demonstrate stability of the \emph{%
entire family},\emph{\ }despite the fact that collapse occurs in the 1D
equation with the quintic self-focusing term. In Section III, we investigate
soliton collisions within the framework of this equation. On the contrary to
completely elastic collisions between solitons in the cubic GPE (alias cubic
nonlinear Schr\"{o}dinger equation, NLSE), in the presence of the quintic
term colliding solitons with zero phase difference, $\Delta \varphi =0$,
merge into a single pulse if their relative velocity is smaller than a
critical value, $2V_{c}$. We find the dependence of $V_{c}$ on the strength
of the transverse confinement and number of atoms in the solitons. For
moderate quintic nonlinearity, good agreement with an analytic prediction
derived from the perturbation theory is found. With a stronger quintic term,
the numerical results deviate from the perturbation theory, although not
dramatically. Finally, we demonstrate dynamical symmetry breaking between
identical solitons colliding with $\triangle \varphi \neq 0$ (in that case,
the merger does not occur), as a function of the relative velocity. An
explanation to the latter effect is proposed. It is based on estimation of a
symmetry-breaking parameter, which is a mismatch between the \textit{%
amplitude center} and \textit{phase center} of the soliton pair with $\Delta
\varphi \neq 0$ (exact definitions are given below). Reasonably good
agreement between numerical results and the analytical approximation is
observed. The paper is concluded by Section IV.

\section{An effective one-dimensional Gross-Pitaevskii equation and exact
soliton solutions}

\subsection{Basic equations}

We start with the standard GPE for a condensate tightly confined in the
transverse plane, with the radial coordinate $r$, and unconfined in the
axial direction, $x$:
\begin{equation}
i\hbar \frac{\partial \psi }{\partial t}=-\frac{\hbar ^{2}}{2m}\left( \nabla
_{\perp }^{2}+\frac{\partial ^{2}}{\partial x^{2}}\right) \psi +\frac{1}{2}%
m\omega ^{2}r^{2}\psi +\frac{4\pi \hbar ^{2}a}{m}|\psi |^{2}\psi ,
\label{GPE}
\end{equation}
where operator $\nabla _{\perp }^{2}$ acts in the transverse plane, $\omega $
is the frequency of the trapping potential in this plane, $m$ is the atomic
mass, and $a<0$ is the scattering length. Transition to the quasi-1D
description is possible if the change of the chemical potential due to the
mean-field interaction is much smaller than the level spacing in the
transverse trapping potential. We briefly recapitulate the corresponding
derivation, following, chiefly, Ref. \cite{Brand06}. In the quasi-1D limit,
the factorized ansatz, $\psi (r,x,t)=\phi (x,t)\chi (r,x,t)$ \cite%
{PerezGarcia98}, is used to adiabatically separate fast transverse and slow
longitudinal dynamics, by neglecting derivatives of $\chi $ with respect to
the slow variables, $x$ and $t$. By substituting the ansatz into Eq. (\ref%
{GPE}), two decoupled equations are obtained, within the framework of the
tight-transverse-confinement approximation:
\begin{equation}
i\hbar \frac{\partial \phi }{\partial t}=-\frac{\hbar ^{2}}{2m}\frac{%
\partial ^{2}\phi }{\partial ^{2}x}+\tilde{\mu}\phi,  \label{GPElong}
\end{equation}

\begin{equation}
\tilde{\mu}\chi =-\frac{\hbar ^{2}}{2m}\nabla _{\bot }^{2}\chi +\frac{1}{2}%
m\omega ^{2}r^{2}\chi +\frac{4\pi \hbar ^{2}a}{m}n|\chi |^{2}\chi ,
\label{GPEtransv}
\end{equation}%
where the transverse chemical potential, $\tilde{\mu}$, has to be found from
the ground-state solution of Eq. (\ref{GPEtransv}) as a function of the 1D
density, $n(x,t)\equiv |\phi (x,t)|^{2}$. Physical solutions of Eq. (\ref%
{GPEtransv}) exist only if $-an<0.47$ \cite{Kivshar00}, otherwise transverse
collapse occurs \cite{Brand04}. In the quasi-1D limit, corresponding to $%
-an<<0.47$, the transverse wave function, $\chi $, is close to the ground
state of the 2D harmonic potential, and can be expanded over the set of
transverse eigenmodes, $\varphi _{m}(r)$: $\chi (r,x)=\varphi _{0}(r)+\Sigma
_{m}C_{m}(x)\varphi _{m}(r)$. Coefficients $C_{m}$ are small and can be
calculated perturbatively. Accordingly, the transverse chemical potential $%
\tilde{\mu}$ can be expanded over powers of the density by means of the
perturbative theory, $\tilde{\mu}=\hbar \omega +g_{\mathrm{1D}%
}n-g_{2}n^{2}+...$, where
\begin{equation}
g_{\mathrm{1D}}=2\hbar \omega a,~g_{2}=24\left( \ln \frac{4}{3}\right) \hbar
\omega a^{2},  \label{gg}
\end{equation}%
as shown in Ref. \cite{Shlyap02} (the subscript ``1D" implies that
the corresponding coefficient appertains to the standard 1D
model). Substituting the expansion for $\tilde{\mu}$ in Eq.
(\ref{GPElong}), one arrives at an effective equation describing
the soliton dynamics in the quasi-1D limit:
\begin{equation}
i\hbar \frac{\partial \phi }{\partial t}=-\frac{\hbar ^{2}}{2m}\frac{%
\partial ^{2}\phi }{\partial ^{2}x}+g_{\mathrm{1D}}|\phi |^{2}\phi
-g_{2}|\phi |^{4}\phi ,  \label{GPEquintic}
\end{equation}%
which is NLSE with the cubic-quintic (CQ) nonlinearity.

Other approaches to the derivation of the effective 1D GPE were also
proposed \cite{Salasnich,Isaac}. In particular, a more complex equation with
nonpolynomial (algebraic) nonlinearity was derived, by means of the
variational approach to the separation of the axial and transverse wave
functions, in Ref. \cite{Salasnich}. Expanding the nonlinearity up to the
quintic term, one arrives at an equation similar to Eq. (\ref{GPEquintic}),
but with a different numerical coefficient.

NLSEs with the CQ nonlinearity are well known as model equations in
nonlinear optics, starting with pioneer works \cite{Pushkarov79}. GPEs with
the CQ nonlinearity were also used in order to take into account three-body
collisions in the BEC \cite{Fatkhulla}. However, in the previously
considered settings, these equations were always considered with a
combination of self-focusing cubic and \emph{self-defocusing} quintic terms.
A drastic difference in the present case is that the quintic term is \emph{%
self-focusing} [as seen from Eq. (\ref{gg}), this conclusion does not depend
on the sign of scattering length $a$, i.e., on the self-focusing or
defocusing character of the cubic term; the same conclusion follows from the
expansion of the above-mentioned nonpolynomial NLSE derived in Ref. \cite%
{Salasnich}]. The use of the GPE with the ``double-self-focusing"
CQ nonlinearity, which is the case here, was tacitly assumed
impossible, as in this case the equation gives rise to collapse.
Nevertheless, we will show below that this equation generates
meaningful stable solutions. In fact, if the cubic nonlinearity is
self-focusing, i.e., the scattering length is negative (the case
considered throughout the present work), the presence of the
collapse is a relevant qualitative feature of the effective GPE,
as collapse takes place too in the full 3D
equation, from which Eq. (\ref{GPEquintic}) was derived (even if the \textit{%
strong collapse} in the full 3D GPE and \textit{weak collapse} in the 1D CQ
equation bear essential differences). As shown in Refs. \cite%
{PerezGarcia98,Carr02,Gammal02}, the collapse in the 3D equation may be
avoided under the constraint of $N|a|/a_{\perp }<0.627$, where $N$ is the
number of atoms in the condensate, and $a_{\perp }=\sqrt{\hbar /(m\omega )}$
is the harmonic-oscillator length corresponding to the transverse
confinement.

\subsection{Soliton solutions}

Below, we use Eq. (\ref{GPEquintic}) in the normalized form,
\begin{equation}
i\frac{\partial \phi }{\partial t}=-\frac{1}{2}\frac{\partial ^{2}\phi }{%
\partial ^{2}x}+g_{\mathrm{1D}}|\phi |^{2}\phi -g_{2}|\phi |^{4}\phi ,
\label{GPEdless}
\end{equation}%
where $g_{\mathrm{1D}}<0$ and $g_{2}>0$ are dimensionless interaction
constants. In fact, the absolute values of both of them may be additionally
scaled to be $1$, but we find it more convenient to keep these coefficients
as free parameters.

\begin{figure}[tbp]
{\centering \resizebox*{0.5\textwidth}{0.3\textheight} {{%
\includegraphics{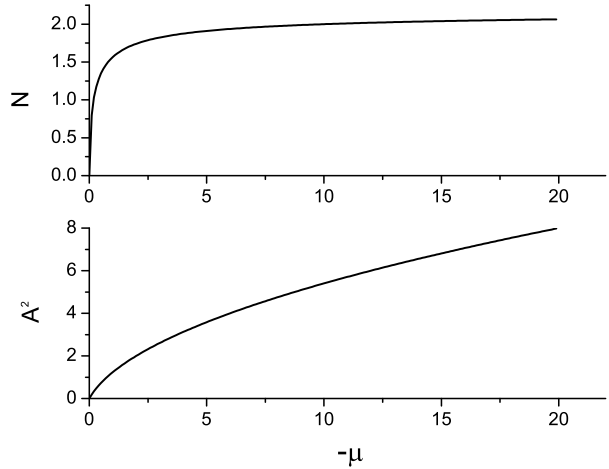}}} }
\caption{The upper and lower panels display, respectively, the family of
exact soliton solutions (\protect\ref{CQsolution}), with $g_{\mathrm{1D}}=-1$
and $g_{2}=3/4$ [hence, $g=1$, see Eq. (\protect\ref{g})], in terms of the
dependences of the norm and squared amplitude vs the chemical potential, as
per Eqs. (\protect\ref{NormCQ}) and (\protect\ref{AmplCQ}).}
\label{OneSolitonSolution}
\end{figure}

A family of exact soliton solutions to Eq. (\ref{GPEdless}) can be found as
an analytical continuation of the well-known solution of the equation with
the self-defocusing quintic term \cite{Pushkarov79}. The result is
\begin{eqnarray}
\phi (x,t) &=&2\left( \frac{3}{4g_{2}}\right) ^{1/4}e^{-i\mu t}\sqrt{\frac{%
-\mu }{\sqrt{g^{2}-4\mu }\cosh \left( 2\sqrt{-2\mu }x\right) +g}}~,
\label{CQsolution} \\
g &\equiv &-\frac{1}{2}\sqrt{\frac{3}{g_{2}}}g_{\mathrm{1D}}~,  \label{g}
\end{eqnarray}%
where $\mu $ is the soliton's chemical potential that may take any value
from $0<-\mu <\infty $. The squared amplitude of this soliton, i.e., the
maximum atomic density at its center, is
\begin{equation}
A^{2}=\frac{1}{2}\sqrt{\frac{3}{g_{2}}}\left( \sqrt{g^{2}-4\mu }-g\right) ,
\label{AmplCQ}
\end{equation}%
and the norm of the soliton, which measures the total number of atoms, is
\begin{equation}
N_{\mathrm{sol}}\equiv \int_{-\infty }^{+\infty }|\phi (x)|^{2}dx=\sqrt{%
\frac{6}{g_{2}}}\tan ^{-1}\left( \frac{2\sqrt{-\mu }}{\sqrt{g^{2}-4\mu }+g}%
\right) ~.  \label{NormCQ}
\end{equation}%
The soliton's norm and squared amplitude are shown, as functions of the
chemical potential, in Fig. \ref{OneSolitonSolution}.

It is worth to note a drastic difference of this soliton family from its
counterpart in the model with the self-defocusing quintic term, i.e., $%
g_{2}<0$: in that case, the norm takes all values, $0<N_{\mathrm{sol}%
}<\infty $, while the chemical potential is limited to a finite interval, $%
0<-\mu <|\mu |_{\max }\equiv 3g_{\mathrm{1D}}^{2}/\left( 16\left\vert
g_{2}\right\vert \right) $. On the contrary, for the present solution
family, Eq. (\ref{NormCQ}) demonstrates that the norm is limited to a finite
interval,
\begin{equation}
0<N_{\mathrm{sol}}<N_{\max }=\sqrt{\frac{3}{8g_{2}}}\pi ,  \label{max}
\end{equation}%
while $-\mu $ is not limited from above (as said above). In fact, $N_{\max }$
in Eq. (\ref{max}) is a collapse border of the soliton family. Further, in
the usual CQ model, with $g_{2}<0$, the amplitude is limited by a finite
value, $A^{2}<3g_{\mathrm{1D}}/\left( 4g_{2}\right) $, while the width of
the soliton diverges $\sim \ln \left( \left( |\mu |_{\max }-|\mu |\right)
^{-1}\right) $ at $|\mu |\rightarrow |\mu |_{\max }$. In contrast to this,
Eqs. (\ref{CQsolution}) and (\ref{AmplCQ}) show that the amplitude of the
present soliton family diverges, $A^{2}\approx \sqrt{-\left( 3/g_{2}\right)
\mu }$, at $\mu \rightarrow -\infty $, and the width of the soliton shrinks
in the same limit, as $1/\sqrt{-\mu }$. This asymptotic behavior of the
soliton solution clearly suggests a transition to a collapsing solution at $%
N_{\mathrm{sol}}\rightarrow N_{\max }$, see Eq. (\ref{max}).

Equation (\ref{NormCQ}) shows that condition $dN/d\mu <0$ holds for the
entire soliton family (see also the upper panel in Fig. \ref%
{OneSolitonSolution}), hence the solitons satisfy the known \textit{%
Vakhitov-Kolokolov }(VK)\textit{\ }stability criterion \cite{VK}. As this
criterion is only a necessary one, but not sufficient, the stability of the
solitons was tested in systematic direct simulations of Eq. (\ref{GPEdless}%
). Results clearly suggest that \emph{all} the solitons are indeed stable
against small perturbations (of course, a large perturbation may provoke
onset of the collapse).

It is relevant to mention that the 1D GPE with the full algebraic
nonlinearity introduced in Ref. \cite{Salasnich} gives rise to \emph{two
branches} of (implicit) soliton solutions, one stable and one unstable; the
branches meet and disappear at the point of transition to collapsing
solutions. Equation (\ref{GPEdless}) does not give rise to the second
branch, as the combination of the cubic and quintic terms may be regarded as
a truncated expansion of the full algebraic nonlinearity from the
above-mentioned equation, and this truncation does not pick up the unstable
branch.

It may also be relevant to note that, starting the derivation of the
effective 1D equation from the 3D\ GPE with the positive scattering length
(corresponding to self-repulsive BEC), one will arrive at Eq. (\ref{GPEdless}%
) with $g_{\mathrm{1D}}>0$ (and again with $g_{2}>0$). The corresponding
equation, featuring competition between the cubic self-focusing and quintic
self-defocusing terms, has a family of exact soliton solutions given by the
same expressions (\ref{CQsolution})-(\ref{NormCQ}), in which $g$ is
negative, as per Eq. (\ref{g}). Despite the formal similarity, the latter
soliton family is completely different from the one presented above. In
particular, in the limit of $\mu \rightarrow 0$ the solution is not a usual
broad small-amplitude soliton, but rather an algebraic one,%
\begin{equation}
\phi _{\mu =0}(x)=\left( \frac{3}{g_{2}}\right) ^{1/4}\sqrt{\frac{-g}{%
1+2g^{2}x^{2}}}.  \label{mu=0}
\end{equation}%
The most drastic difference of the soliton family with $g_{\mathrm{1D}}>0$
and $g_{2}>0$ from the above one is that it features $dN/d\mu >0$, hence
this \emph{entire family} is unstable, according to the VK criterion
[algebraic solitons, such as one in Eq. (\ref{mu=0}), are known to be
unstable for a different reason \cite{Seva}]. Besides the fact that all the
solitons in the model with the positive scattering length are unstable,
their physical meaning is doubtful also because the quintic term, which
appears as a perturbative correction to the cubic one \cite{Shlyap02},
actually dominates over it in these solutions.

\section{Soliton collisions}

\subsection{Merger of colliding solitons with $\Delta \protect\varphi =0$}

It is commonly known that collisions between solitons in the one-dimensional
NLSE, which is an integrable equation, are completely elastic. The force of
interaction between the solitons depends on the relative phase between them:
with $\triangle \varphi =0$ and $\Delta \varphi =\pi $, they are attract and
repel each other, respectively \cite{KarpmanSol,Malomed89}. The quintic term
breaks the integrability of the equation, and is expected to make collisions
inelastic. For $\triangle \varphi =0$, simulations reveal a critical
collision velocity, below which two identical solitons merge into a single
one.

\subsubsection{Analytical considerations}

The merger may be explained by the fact that radiation loss due to the
inelastic collision becomes greater than the initial kinetic energy of the
soliton pair \cite{Malomed89}. This explanation can be implemented in an
explicit form if the quintic term is treated as a small perturbation. To
this end, defining $\Psi \equiv \sqrt{\left\vert g_{\mathrm{1D}}\right\vert }%
\phi $, we rewrite Eq. (\ref{GPEdless}) in the following form:
\begin{equation}
i\frac{\partial \Psi }{\partial t}=-\frac{1}{2}\frac{\partial ^{2}\Psi }{%
\partial ^{2}x}-|\Psi |^{2}\Psi -\epsilon |\Psi |^{4}\Psi ,
\label{GPEperturbed}
\end{equation}
where $\epsilon \equiv g_{2}/g_{\mathrm{1D}}^{2}$. In the zero-order
approximation ($\epsilon =0$), the traveling-soliton solution to Eq. (\ref%
{GPEperturbed}) is
\begin{equation}
\Psi (x,t)=A\ \mathrm{sech}(A(x-Vt))e^{-i(\mu t-Vx)},
\label{Qtravelsolution}
\end{equation}
where $A$ and $V$ are its amplitude and velocity, and the frequency $\omega
=V^{2}/2+\mu $ is a sum of the kinetic energy and binding (potential)
energy, $\mu =-A^{2}/2$, per particle.

The use of the perturbation theory makes it possible to obtain the following
analytical result for the collision between solitons with equal amplitudes $%
A $, velocities $\pm V$, and a phase shift $\triangle \varphi _{0}$ between
them \cite{Malomed89}: if the solitons are fast, $V^{2}\gg A^{2}$, the
energy loss generated by the emission of radiation during the collision is
\begin{equation}
(\triangle E)_{\mathrm{rad}}=\epsilon ^{2}A^{2}\left\{ \alpha
A^{5}+V^{5}e^{-\pi V/A}[\beta _{1}\cos (\triangle \varphi )+\beta _{2}\sin
(\triangle \varphi )]\right\} ,  \label{Energyloss}
\end{equation}
where $\alpha \approx 1381$,$\beta _{1}\approx 2401$ and $\beta _{2}\approx
347$. Note that the phase-dependent terms are exponentially small. In the
same approximation, the collision-induced loss of the number of atoms is
\begin{equation}
(\triangle N)_{\mathrm{rad}}=\left( 2/V^{2}\right) (\triangle E)_{\mathrm{rad%
}}~.  \label{Numberloss}
\end{equation}

To estimate a merger condition (threshold), we assume that the velocities $%
\pm V$, which determine the collision-induced losses as per Eqs. (\ref%
{Energyloss}) and (\ref{Numberloss}), are actually acquired by originally
quiescent (or slowly moving) solitons due to their mutual attraction (if $%
\triangle \varphi $ is close to zero). To this purpose, we note that the
effective potential of the interaction between far separated identical
solitons is, in the case of $\epsilon =0$,
\begin{equation}
U_{\mathrm{int}}(X,\triangle \varphi _{0})=-8A^{3}e^{-AX}\cos (\triangle
\varphi _{0})  \label{Potential}
\end{equation}%
\cite{KarpmanSol}, and the effective mass of the soliton is $M_{\mathrm{eff}%
}=2A$. In this approximation, the attraction accelerates the two in-phase
solitons to \textit{self-acquired velocities}, $\pm V_{\mathrm{self}}$, that
can be found from the energy-balance equation, $2\cdot \left( M_{\mathrm{eff}%
}V_{\mathrm{self}}^{2}/2\right) =8A^{3}$, hence $V_{\mathrm{self}}=2A$.
Substituting this velocity in Eq. (\ref{Energyloss}) shows that the
phase-dependent part is less than $10\%$ of the phase-independent one, and
therefore we neglect it. Thus, the collision-induced loss of the energy and
number of atoms (for both solitons) are predicted by the perturbation theory
to be
\begin{equation}
(\triangle E)_{\mathrm{rad}}=\alpha \epsilon ^{2}A^{7},(\triangle N)_{%
\mathrm{rad}}=(\alpha /2)\epsilon ^{2}A^{5},  \label{collisionloss}
\end{equation}%
where $\alpha $ is the same numerical coefficient as in Eq. (\ref{Energyloss}%
).

The energy of a free soliton and its norm (number of atoms), in the $%
\epsilon =0$ limit, are
\begin{equation}
E_{\mathrm{sol}}=-\frac{1}{3}A^{3}+\frac{1}{2}M_{\mathrm{eff}}V^{2},~N_{%
\mathrm{sol}}=2A  \label{SolitonEnergy}
\end{equation}
(the negative term in $E_{\mathrm{sol}}$ is the binding energy). First, the
norm loss, $\triangle N$, taken from Eq. (\ref{collisionloss}), gives rise
to the collision-induced change of the soliton's amplitude: $\triangle
A=-(\triangle N)_{\mathrm{rad}}/2=-(\alpha /4)\epsilon ^{2}A^{5}$. The
corresponding change in the binding (potential) energy of both solitons is
positive,
\begin{equation}
\triangle E_{\mathrm{bind}}\equiv \triangle (-\frac{2}{3}A^{3})=-2A^{2}%
\triangle A=\frac{\alpha }{2}\epsilon ^{2}A^{7}.  \label{PotentialEChange}
\end{equation}
Finally, the energy balance predicts a change in the total kinetic energy:
\begin{equation}
\triangle E_{\mathrm{kin}}=-(\triangle E)_{\mathrm{rad}}-\triangle E_{%
\mathrm{bind}}=-(3\alpha /2)\epsilon ^{2}A^{7}.  \label{KineticEChange}
\end{equation}

The merger condition states that the loss of the kinetic energy is equal to
or exceeds the initial kinetic energy \cite{Malomed89}. With regard to the
expression for the total kinetic energy of both solitons which follows from
Eq. (\ref{SolitonEnergy}), $E_{\mathrm{kin}}=2AV^{2}$, this condition means
that the merger is expected if the initial velocity of each soliton falls
below a critical value:
\begin{equation}
V^{2}<V_{c}^{2}=\frac{3}{4}\alpha \epsilon ^{2}A^{6}\equiv \frac{3\alpha }{%
256}\epsilon ^{2}N_{\mathrm{sol}}^{6}~.  \label{CriticalVelocity}
\end{equation}%
The derivation of the merger threshold implies that the critical velocity is
much smaller than the above-mentioned self-acquired velocity, $V_{\mathrm{%
self}}=2A$ (then, the initial velocities of the solitons may be disregarded
in the above energy-balance analysis, in comparison with $V_{\mathrm{self}}$%
, as it was actually done). Expression (\ref{CriticalVelocity}) indeed
satisfies condition $V_{c}\ll V_{\mathrm{self}}$, as $\epsilon $ is a small
parameter.

\subsubsection{Numerical results}

For simulations of soliton collisions in Eq. (\ref{GPEdless}), we chose
parameter values close to those in the real experiment \cite{Khaykovich02},
where $^{7}$Li atoms were used: a very small scattering length, $a=-0.06$
nm, transverse oscillation frequency $\omega =2\pi \times 710$ Hz, and the
number of atoms $N_{\mathrm{sol}}=4000$. However, we did not include any
external longitudinal potential, in contrast to the expulsive potential that
was present in the experiment. Recall that the expulsive potential made the
soliton stability region very small \cite{Khaykovich02}, and actually caused
the soliton to be very close to the 3D limit. The present simulations do not
include the external potential because we are interested not in effects
produced by such a potential, but rather in small deviations from the
one-dimensionality. In fact, a modification of the above-mentioned
experimental setup, with the aim to make the central segment of the
cigar-shaped trap free of any tangible axial potential, is quite possible.

\begin{figure}[tbp]
{\centering \resizebox*{0.5\textwidth}{0.3\textheight} {{%
\includegraphics{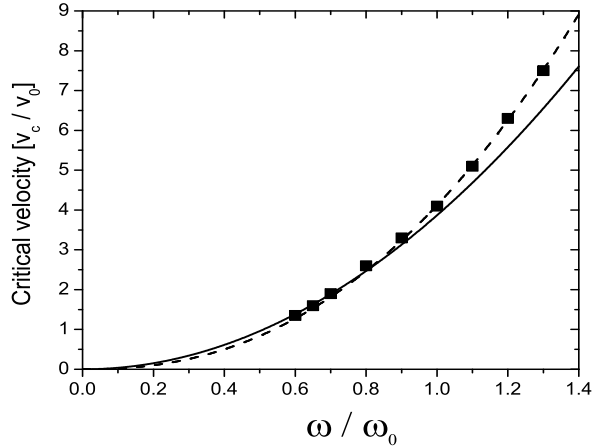}}} }
\caption{The critical velocity for the merger of colliding solitons, $%
v_{c}/v_{0}$ (as an experimentally relevant reference value, we take $%
v_{0}=0.21$ mm/s), as a function of the strength of the transverse
confinement, $\protect\omega /\protect\omega _{0}$ (with $\protect\omega %
_{0}=2\protect\pi \times 710$ Hz, as in Ref. \protect\cite{Khaykovich02}).
For relatively weak confinement (smaller $\protect\omega $), dependence $%
V_{c}\propto \protect\omega ^{2}$ is observed, as predicted by the
perturbation theory (the solid line shows the $\protect\omega ^{2}$ power
law as a guide to the eye). The dashed line is the power-law fit to the last
four points of the numerical results, yielding $V_{c}\propto \protect\omega %
^{2.29\pm 0.07}$.}
\label{VcEpsilon}
\end{figure}

To compare the analytical prediction for the critical velocity, given by Eq.
(\ref{CriticalVelocity}), to numerical results, it is necessary to express
perturbative parameter $\epsilon $ in terms of the transverse trapping
frequency $\omega $. Undoing the above renormalizations, one arrives at a
conclusion that Eq. (\ref{CriticalVelocity}) implies a quadratic dependence,
$V_{c}\propto \omega ^{2}$, within the framework of the perturbation theory.
This dependence is indeed observed in simulations at relatively small $%
\omega $, as seen in Fig. (\ref{VcEpsilon}). However, at larger $\omega $,
i.e., for stronger transverse confinement, the numerical results feature a
greater power in the $V_{c}(\omega )$ dependence. In particular, the best
fit to the last four numerical points in Fig. (\ref{VcEpsilon}) yields $%
V_{c}\propto \omega ^{2.29\pm 0.07}$, which demonstrates a small but
tangible deviation from the power law corresponding to the perturbation
limit.

\begin{figure}[tbp]
{\centering \resizebox*{0.5\textwidth}{0.3\textheight} {{%
\includegraphics{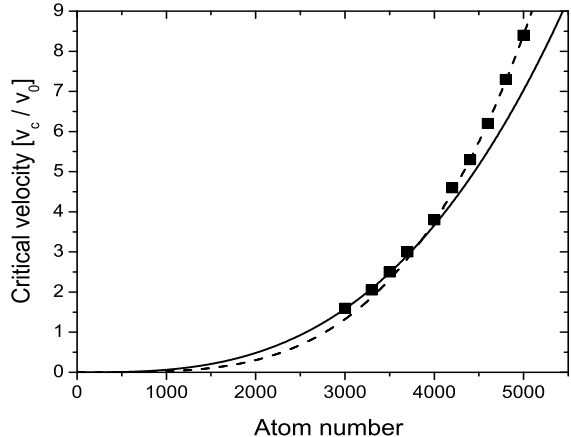}}} }
\caption{The critical velocity for the merger of two solitons as a function
of the number of atoms in each of the colliding solitons, $N_{\mathrm{sol}}$%
. At smaller $N_{\mathrm{sol}}$, i.e., for weaker nonlinearity, the $%
V_{c}\propto N_{\mathrm{sol}}^{3}$ dependence is observed, as predicted by
the perturbation theory, see Eq. (\protect\ref{CriticalVelocity}). To say
more accurately, the solid curve, which is the fit to the first four
numerical points, features a power law $V_{c}\propto N_{\mathrm{sol}%
}^{2.9\pm 0.2}$. The dashed curve is the power-law fit to the last four
points, showing a different power dependence, $V_{c}\propto N_{\mathrm{sol}%
}^{3.62\pm 0.05}$.}
\label{VcNumber}
\end{figure}

\begin{figure}[tbp]
{\centering \resizebox*{0.5\textwidth}{0.3\textheight} {{%
\includegraphics{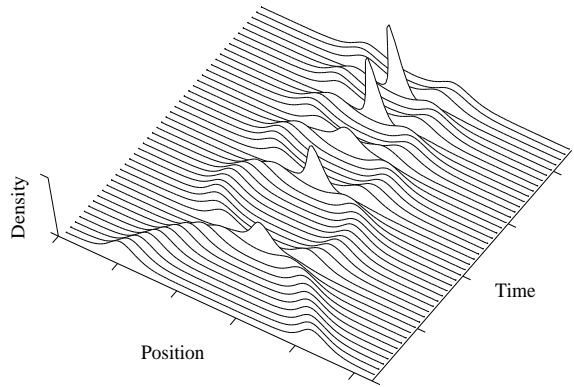}}} }
\caption{Density profiles as a function of time in a regime of
``weak" merger ($N_{\mathrm{sol}}=3500$, $v/v_{0}=2$). After the
first collision, the two solitons re-appear as two distinct wave
pulses which then collide again many times.}
\label{WeakMergerProfiles}
\end{figure}

We now turn to the dependence of the critical velocity on the
number of atoms, $N$. The analytical prediction, Eq.
(\ref{CriticalVelocity}), clearly implies $V_{c}\propto N^{3}$. In
Fig. (\ref{VcNumber}), this dependence is indeed observed at
smaller values of $N$ (i.e., for weaker nonlinearity), where the
perturbation limit should naturally be valid. Perusal of numerical
data shows that, in this range, the actual collision-induced
radiation loss is very small, and, as a result, the merger does
not lead to complete fusion of the colliding solitons into a
single pulse, but rather to formation of a bound state of two
solitons (``weak merger"), as can be seen in Fig.
\ref{WeakMergerProfiles}. Namely, after the first collision, the
solitons re-emerge as two distinct wave packets which then collide
again many times. A similar nearly radiationless inelastic
collision, leading to the formation of a two-soliton loosely bound
state, was recently observed in simulations of a weakly discrete
cubic NLSE \cite{Dmitriev02}.

A definite deviation from the $V_{c}\propto N_{\mathrm{sol}}^{3}$
dependence is observed in Fig. \ref{VcNumber} for
$N_{\mathrm{sol}}>4000$, which shows a limitation of the
perturbative predictions. In this regime of strong nonlinearity, a
smooth transition in the collision process occurs, from the
formation of the above-mentioned long-lived bound state to direct
(``strong") merger of two solitons into a single pulse, which is
accompanied by a burst of radiation. The conspicuous loss of
matter with the radiation prevents the emerging single pulse from
having the number of atoms above the collapse threshold, therefore
the pulse \emph{does not} blow up. The transition is expressed in
reduction of the life time of the
loose bound state before the complete merger. In Fig. \ref%
{StrongMergerProfiles}, which represents the strongest nonlinearity included
in the present framework, the bound state features only two oscillations.

\begin{figure}[tbp]
{\centering \resizebox*{0.5\textwidth}{0.3\textheight} {{%
\includegraphics{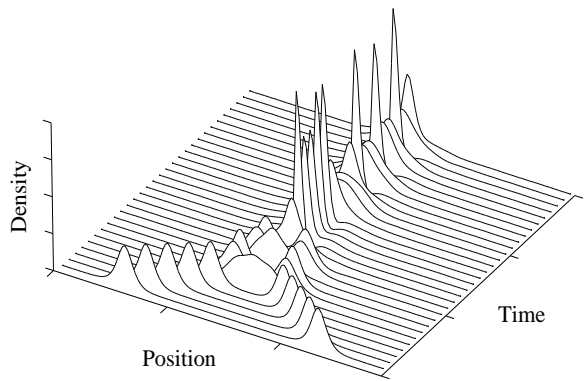}}} }
\caption{The same as in Fig. \protect\ref{WeakMergerProfiles}, but in the
regime of ``strong" merger (for $N_{\mathrm{sol}}=5000$, $%
v/v_{0}=8$). A short-lived bound state quickly merges into a single
breather-like pulse.}
\label{StrongMergerProfiles}
\end{figure}

For even stronger nonlinearities (which were also considered), the power-law
dependence of $V_{c}$ on the number of atoms and transverse trapping
frequency is observed in the form of $N_{\mathrm{sol}}^{4}$ and $\omega ^{3}$%
, respectively. However, in such an extreme regime, the relevance of the
quasi-1D model is questionable. In any case, these results convey a clear
caveat to soliton experiments, in which relative variations in the atom
number may be as large as $\simeq 2$: the strong power-law dependence of the
critical velocity on the number of atoms should be taken into account, to
avoid occasional merger of solitons.

\subsection{Symmetry breaking in soliton collisions with $\Delta \protect%
\varphi \neq 0$}

We proceed to inelastic collisions of identical solitons with the phase
difference of $0<\triangle \varphi <\pi /2$. Numerical simulations of Eq. (%
\ref{GPEdless}) show a salient effect of symmetry breaking in this case:
while the solitons separate after the collision, they emerge as two pulses
with \emph{different amplitudes} (then, the velocities are also different,
to comply with the momentum conservation), as shown in Fig. \ref%
{AsymmetryProfiles}. It should be mentioned that a similar effect was
observed in simulations of collisions between identical solitons in some
other nonintegrable 1D models, chiefly in those describing transmission of
nonlinear optical pulses, within the framework of the coupled-mode theory,
in waveguides equipped with Bragg gratings. In that context, the
collision-induced symmetry breaking was reported in basic single-core models
\cite{SSB-single-core}, and in more sophisticated dual-core ones \cite%
{SSB-dual-core}. A similar effect was also observed in collisions between
moving solitons in the discrete NLSE \cite{Frantzeskakis03}.

\begin{figure}[tbp]
{\centering \resizebox*{0.5\textwidth}{0.3\textheight} {{%
\includegraphics{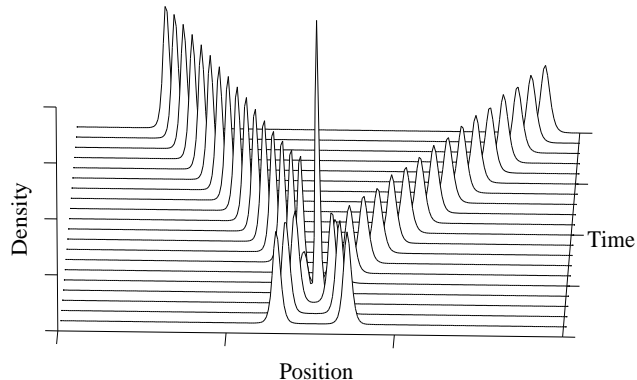}}} }
\caption{Density profiles of solitons featuring the symmetry breaking in the
collision, for $\triangle \protect\varphi =\protect\pi /10$.}
\label{AsymmetryProfiles}
\end{figure}

In order to achieve qualitative understanding of the symmetry breaking, we
resort to consideration of an \textit{ansatz} based on a formal linear
superposition of two completely overlapping identical solitons (\ref%
{CQsolution}), at some moment of time $t=t_{0}$, with velocities $\pm V$ and
phases $\pm (1/2)\triangle \varphi $. The ansatz\textit{\ }yields the
following expression:
\begin{equation}
\Psi _{\mathrm{ansatz}}(x,t)=2e^{i(|\mu |-\frac{1}{2}V^{2})t_{0}}\sqrt{\frac{%
|\mu |}{\sqrt{g^{2}+4|\mu |}\cosh \left( 2\sqrt{2|\mu |}x\right) +g}}\cos
(Vx+\triangle \varphi )~.  \label{CQ2ssolution}
\end{equation}%
An essential peculiarity of this expression is that the central points of
the two last multipliers do not coincide: one is found at $x=0$, while the
other one at $x=-\triangle \varphi /V$. This simple observation suggests a
concept of the mismatch between the \textit{amplitude center} and \textit{%
phase center} of the pair of colliding identical solitons. The mismatch was
considered as a cause of breaking the symmetry between colliding solitons in
the above-mentioned model based on the discrete NLSE \cite{Frantzeskakis03}.

To characterize the asymmetry of ansatz (\ref{CQ2ssolution}) qualitatively,
we introduce its center-of-mass coordinate,
\begin{equation}
\xi (v)\equiv \frac{\int_{-\infty }^{+\infty }x|\phi _{\mathrm{ansatz}%
}(x)|^{2}dx}{\int_{-\infty }^{+\infty }|\phi _{\mathrm{ansatz}}(x)|^{2}dx}=%
\frac{\sin (2\triangle \varphi )}{\sqrt{2|\mu |}}\frac{\gamma \cosh (\gamma
\upsilon )\sinh (\pi \upsilon )-\pi \sinh (\gamma \upsilon )\cosh (\pi
\upsilon )}{\sinh (\pi \upsilon )[\gamma \sinh (\pi \upsilon )+\pi \cos
(2\triangle \varphi )\sinh (\gamma \upsilon )]},  \label{CoM}
\end{equation}%
with $\upsilon \equiv V/\sqrt{2|\mu |}$ and $\gamma \equiv \tan ^{-1}(2\sqrt{%
|\mu |}/g)$. For the qualitative understanding of the situation, we adopt a
natural conjecture that the strongest possible symmetry breaking is attained
at a value of the velocity $\upsilon =\upsilon _{\max }$, which corresponds
to a maximum of $|\xi |$ for given $\triangle \varphi $. For the weak
quintic nonlinearity ($|\mu |\ll g^{2}$), one has $\gamma \approx 2\sqrt{%
|\mu |}/g$, and Eq. (\ref{CoM}) simplifies:
\begin{equation}
\xi (V)\approx \frac{\sin (2\triangle \varphi )}{\sqrt{2|\mu |}}\frac{\sinh
(\pi \upsilon )-\pi \upsilon \cosh (\pi \upsilon )}{\sinh (\pi \upsilon
)[\sinh (\pi \upsilon )+\pi \upsilon \cos (2\triangle \varphi )]},
\label{CenterOfMassWeakNL}
\end{equation}

Asymmetry parameter $\xi \ $is shown, as a function of $V$, in Fig. (\ref%
{AmplitudeRatioPi10}) by the solid line for $\triangle \varphi =\pi /10$. It
characterizes the degree of the collision-induced symmetry breaking, and
predicts a maximum at some \emph{nonzero} velocity. Quite a similar
dependence is indeed produced by numerical simulations of Eq. (\ref{GPEdless}%
) for the same value of $\triangle \varphi $, as shown by dots in Fig. (\ref%
{AmplitudeRatioPi10}). The dots display values of the amplitude ratio of the
output soliton pair, as found from the simulations. Actually, the pulses
emerging from the inelastic collisions are breathers with time-dependent
amplitudes. Therefore, we averaged the amplitudes over long propagation
distances after the collision.

\begin{figure}[tbp]
{\centering \resizebox*{0.5\textwidth}{0.3\textheight} {{%
\includegraphics{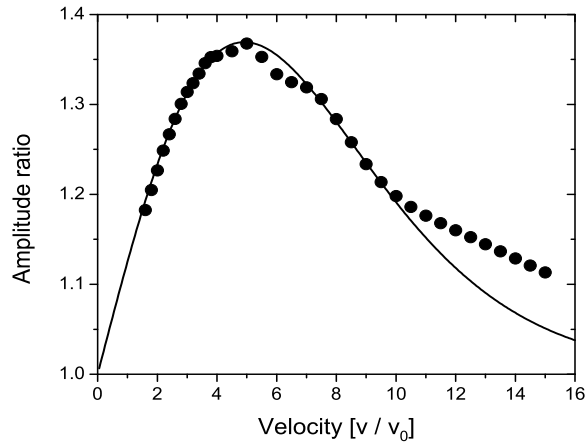}}} }
\caption{Black circles show the numerically found amplitude ratio of two
solitons after the symmetry-breaking collision for initial phase difference $%
\Delta \protect\varphi =\protect\pi /10$. The solid line shows the velocity
dependence of the symmetry-breaking parameter, as given by Eq. (\protect\ref%
{CenterOfMassWeakNL}), for the same value of $\Delta \protect\varphi $.}
\label{AmplitudeRatioPi10}
\end{figure}

Generally, the numerical data in Fig. (\ref{AmplitudeRatioPi10}) follow the
predicted symmetry-breaking parameter. However, two notable deviations are
observed: a nonsmooth shape of the numerical curve (local modulations near
the maximum, which tend to give rise to two extra local maxima, definitely
exceed an error of the numerical simulations), and a weakly decaying tail,
which implies that the asymmetry generated by collisions between fast
solitons is essentially larger than predicted by the above model. Note that
the analytical model does not include radiation loss. The latter may enhance
the asymmetry, as the loss itself is, plausibly, asymmetric too.

We note that Ref. \cite{Adams06}, in which collisions of nearly 3D
solitary waves were considered through simulations of the full 3D
GPE, showed very little symmetry breaking (``population transfer")
between colliding solitons with the initial phase difference of
$\Delta \varphi =\pi /10$, less than $1\%$ . However, our results
predict that the matter transfer (symmetry breaking) would be
conspicuous at specific values of the collision velocities, which
might not be included in the analysis reported in Ref.
\cite{Adams06}

For very small $\Delta \varphi $, we observed chaotic behavior in the output
of the collision, similar to what was reported in a weakly discrete NLSE
\cite{Dmitriev02} (see also Ref. \cite{SalernoModel}). Very recently,
chaotic behavior was predicted for collisions of more than two MW solitons,
in the presence of a longitudinal parabolic trapping potential \cite%
{Martin06}. In our model, the collision between two solitons is sufficient
to observe chaotic behavior, which will be reported elsewhere.

\section{Conclusions}

This work aims to understand how the tight confinement in transverse
directions affects the longitudinal dynamics of matter-wave solitons in the
quasi-1D setting. Within the framework of the known model, which reduces the
multidimensional character of the full Gross-Pitaevskii equation to the
appendage of an additional self-focusing quintic term to the effective 1D
equation, we have investigated deviations from the ideal soliton behavior.

A family of exact stationary solutions for the solitons has been
constructed, and it was demonstrated that the entire family is
stable, despite the possibility of collapse in the modified 1D
equation (with the negative scattering length). We have found
inelastic effects in soliton collisions, which are impossible in
ideal solitons. Two identical in-phase solitons merge into a
single pulse, if the collision velocity is smaller than a critical
value. In fact, two different types of the merger were observed,
``strong" and ``weak" ones, the former leading to the formation of
a loose bound state of two solitons that feature repeated
collisions, with very weak radiation loss, while the latter means
direct fusion into a single pulse, which is accompanied by a burst
of radiation (in that case, the radiation loss helps the emerging
pulse to drop the number of atoms below the collapse threshold,
and thus avoid the blowup). Both the analytical approximation,
based on the perturbation theory, and numerical results highlight
the strong dependence of the critical velocity on the strength of
the transverse confinement and the number of atoms in the
solitons. Symmetry breaking in collision between identical
solitons with nonzero phase difference was also found, and
partially explain by means of the calculation of a
phenomenologically defined symmetry-breaking parameter, which
measures the mismatch between amplitude and phase centers of the
colliding solitons.

This work was supported, in a part, by the Israel Science Foundation,
through grant No. 1125/04 (L.K.) and the Center-of-Excellence grant No.
8006/03 (B.A.M.).

\end{document}